\documentclass[draft]{agujournal2019}
\usepackage{url} %this package should fix any errors with URLs in refs.
\usepackage{soul}
\draftfalse

\journalname{JGR: Planets}

\begin{document}

%% ------------------------------------------------------------------------ %%
%  Title
%
% (A title should be specific, informative, and brief. Use
% abbreviations only if they are defined in the abstract. Titles that
% start with general keywords then specific terms are optimized in
% searches)
%
%% ------------------------------------------------------------------------ %%
\title{Impacts into Titan's methane-clathrate crust as a source of atmospheric methane}

%% ------------------------------------------------------------------------ %%
%
%  AUTHORS AND AFFILIATIONS
%
%% ------------------------------------------------------------------------ %%

% Authors are individuals who have significantly contributed to the
% research and preparation of the article. Group authors are allowed, if
% each author in the group is separately identified in an appendix.)

% List authors by first name or initial followed by last name and
% separated by commas. Use \affil{} to number affiliations, and
% \thanks{} for author notes.
% Additional author notes should be indicated with \thanks{} (for
% example, for current addresses).
\authors{S. Wakita\affil{1,2}, B. C. Johnson\affil{2,3}, J. M. Soderblom\affil{1}, J. K. Steckloff\affil{4,5}, A. V. Johnson\affil{2}, C. D. Neish\affil{4,6}, J. Shah \affil{6}, P. Corlies \affil{7}}

\affiliation{1}{Department of Earth, Atmospheric and Planetary Sciences, Massachusetts Institute of Technology, Cambridge, MA, USA}
\affiliation{2}{Department of Earth, Atmospheric, and Planetary Sciences, Purdue University, West Lafayette, IN, USA}
\affiliation{3}{Department of Physics and Astronomy, Purdue University, West Lafayette, IN, USA}
\affiliation{4}{The Planetary Science Institute, Tucson, AZ, USA}
\affiliation{5}{Department of Aerospace Engineering and Engineering Mechanics, University of Texas at Austin, Austin, TX, USA}
\affiliation{6}{Department of Earth Sciences, The University of Western Ontario, London, ON, Canada}
\affiliation{7}{Spectral Sciences, Burlington, MA, USA}

%% Corresponding Author:
% Corresponding author mailing address and e-mail address:
\correspondingauthor{Shigeru Wakita}{shigeru@mit.edu}
%% Keypoints, final entry on title page.

%  List up to three key points (at least one is required)
%  Key Points summarize the main points and conclusions of the article
%  Each must be 140 characters or fewer with no special characters or punctuation and must be complete sentences
\begin{keypoints}
\item To understand the presence of methane in Titan's atmosphere, we simulate impacts on Titan into a methane-clathrate layer.
\item Impacts release methane gas that enhance its lifetime in Titan's atmosphere.
\item Multiple large impacts might release enough methane to measurably alter Titan's climate.
\end{keypoints}

%% ------------------------------------------------------------------------ %%
%
%  ABSTRACT and PLAIN LANGUAGE SUMMARY
%
% A good Abstract will begin with a short description of the problem
% being addressed, briefly describe the new data or analyses, then
% briefly states the main conclusion(s) and how they are supported and
% uncertainties.

% The Plain Language Summary should be written for a broad audience,
% including journalists and the science-interested public, that will not have 
% a background in your field.
%
% A Plain Language Summary is required in GRL, JGR: Planets, JGR: Biogeosciences,
% JGR: Oceans, G-Cubed, Reviews of Geophysics, and JAMES.
% see http://sharingscience.agu.org/creating-plain-language-summary/)
%
%% ------------------------------------------------------------------------ %%

%% \begin{abstract} starts the second page
%[Be less than 250 words for all journals] 
% https://www.agu.org/publications/authors/journals/text-graphics-requirements#abstract
\begin{abstract}

\end{abstract}

\section*{Plain Language Summary}
Titan is the only icy satellite with a dense atmosphere. 
Methane is the second most abundant molecule in Titan's atmosphere, but sunlight quickly destroys it, so it shouldn't be present there. 
To explain methane's presence in Titan's atmosphere, we may need a mechanism to resupply methane. 
As liquid methane soaks into Titan's icy crust, the ice may trap methane molecules and form methane clathrate. 
This layer would cover Titan’s surface and could serve as a methane reservoir. 
In our work, we simulate cometary impacts into Titan's methane-clathrate layer to determine how much methane is released.
We find that the amount of methane released depends on the impactor size and the thickness of the methane-clathrate layer: a 20-km-diameter impactor releases up to 1\% of the current methane mass in Titan's atmosphere. 
The lifetime of methane would be 3\% longer at the most, if we consider oblique impacts into a porous thick clathrate layer. 
This is not sufficient to explain the presence of methane in Titan's atmosphere. 
However, if multiple large impacts occurred, they would have caused a greenhouse warming effect on Titan.

%% ------------------------------------------------------------------------ %%
%
%  TEXT
%
%% ------------------------------------------------------------------------ %%

%%% Suggested section heads:
% \section{Introduction}
%
% The main text should start with an introduction. Except for short
% manuscripts (such as comments and replies), the text should be divided
% into sections, each with its own heading.

% Headings should be sentence fragments and do not begin with a
% lowercase letter or number. Examples of good headings are:

% \section{Materials and Methods}
% Here is text on Materials and Methods.
%
% \subsection{A descriptive heading about methods}
% More about Methods.
%
% \section{Data} (Or section title might be a descriptive heading about data)
%
% \section{Results} (Or section title might be a descriptive heading about the
% results)
%
% \section{Conclusions}

\section{Introduction}
Titan is the only icy satellite in the solar system that has a dense atmosphere.
While nitrogen is the most abundant molecule in Titan's atmosphere, methane is the second most abundant molecule, with a mixing ratio of $\sim$1.5\% in the stratosphere \cite{Niemann:2005,Horst:2017}. 
The presence of methane in Titan's atmosphere has significant implications for its surface; methane rains onto the surface \cite{Turtle:2011,Turtle:2009} and forms lakes in polar regions \cite{Hayes:2008}. 
It also assumed to form methane clathrate hydrates in the upper ice crust \cite{Choukroun:2010}. 
Experiments in which liquid ethane was mixed with water ice show that they form clathrate on a timescale of 1 year \cite{Vu:2020}. 
If liquid methane behaves in a similar way, we would anticipate that methane clathrate could also form on Titan over short timescales.
However, methane is destroyed in the atmosphere via UV photolysis, producing ethane and heavier organics, which settle on Titan's surface, and hydrogen gas, which escapes to space \cite{Wilson:2009}, making this an irreversible process \cite{Yung:1984,Wilson:2004,Horst:2017}. 
The lifetime of methane in Titan's atmosphere is geologically short.
Although the net loss rate depends on the model, methane in Titan's atmosphere would be entirely lost in 10--100 Myr without resupply \cite{Yung:1984, Wilson:2004}. 
If we assume the current mass of methane is $2\times10^{17}$ kg and the net mass loss rate is $\simeq 300$ kg/s \cite{Tobie:2012,Wilson:2009}, Titan's methane will be lost in $\simeq20$ Myr. 
This begs the question - is Titan's methane-rich atmosphere a geologic anomaly, or is there a mechanism that is resupplying the methane?

Several theories have been proposed to explain the origin of Titan's atmospheric methane.
It may initially be released from Titan's core as a result of dehydration of a hydrous core \cite{Castillo-Rogez:2010,Miller:2019} and subsequently distributed into Titan's ocean and convective ice mantle, forming methane-clathrate deposits \cite{Tobie:2006,Carnahan:2022}. 
Later interior evolution may have destabilized these deposits, releasing methane into Titan's atmosphere \cite{Tobie:2006}.
Another possible source of methane is cryovolcanism \cite{Lopes:2013}. 
Cryovolcanism can also destabilize methane-clathrate deposits, releasing methane into the atmosphere \cite{Davies:2016}.

Similarly, multiple theories have been put forward to explain the resupply of methane to Titan's atmosphere, which requires both a reservoir and a release mechanism.
One such reservoir could be large subsurface pockets of methane \cite<“methane aquifers”,>{Lunine:1983}; such pockets could be in communication with the atmosphere such that the observed surface lakes and seas are only a fraction of the liquid methane cycling between Titan's surface and atmosphere.
Methane clathrates could serve as another reservoir; \citeA{Tobie:2006} proposed that methane clathrate could be stable on Titan's surface.
Subsequent model and experimental works have confirmed its stability \cite{Choukroun:2010,Choukroun:2012}.
Releasing methane from this surface clathrate layer to resupply Titan's atmospheric methane may occur via substitution of ethane for methane in gas hydrate clathrates \cite{Choukroun:2012} or thermal or mechanical destabilization.

Impact cratering can destabilize methane clathrates and/or vaporize liquid methane underground acting as a source of atmospheric methane \cite{Zahnle:2014}. 
To date, however, this process has received little attention. 
To explore the efficiency of impacts to release methane to Titan's atmosphere, we perform simulations of impacts into methane-clathrate layers on Titan using the iSALE-2D shock physics code. 
This differs from previous work that examined the release of methane via impacts, as that work simply estimated the impact energy based on crater scaling laws \cite{Zahnle:2014}. 
In this work, we simulate a range of impactors striking several different thicknesses of methane-clathrate layers and determine how much methane is released by considering the dissociation curve of methane clathrate. 
We then compare our outputs with the current methane mass loss rate and the net loss rate on Titan, and discuss the role of released methane on Titan's atmosphere.

\section{Methods} \label{sec:methods}
We use the iSALE-2D shock physics code to simulate impacts into Titan, assuming it has an ice crust covered by a methane-clathrate layer. 
iSALE is based on the SALE hydrocode \cite{Amsden:1980} and has been developed to model planetary impacts \cite{Collins:2016, Wunnemann:2006}. 
The code has been improved by including various equations of state (EOS) and strength models \cite{Collins:2004, Ivanov:1997, Melosh:1992}.
We simulate spherical ice impactors vertically striking a flat target with an impact velocity of 10.5 km/s, the average impact velocity on Titan \cite{Zahnle:2003}.
Our targets consist of a methane-clathrate layer 5, 10, and 15 km thick overlying water ice. 
Our target thermal profile depends on the clathrate-layer thickness and is given by \citeA{Kalousova:2020}, which indicates a higher near surface thermal gradient due to its low thermal conductivity \cite<about one-fourth that of water ice at 263 K,>{Sloan:2007}.
To model the strength of methane clathrate, we use the same method developed in previous work \cite{Wakita:2022,Wakita:2023}, which considered the material strength of methane clathrate \cite<twenty times stronger than water-ice,>{Durham:2003}.
Neither a shock physics EOS or any Hugoniot data are available for methane clathrate; as such, we use the analytical equation of state (ANEOS) of water ice for methane clathrate. 
Note that high pressure may crush the clathrate cage and release the methane before melting occurs. 
If we can regard this process as the compression within the porous target, it would lead to increasing the internal energy and additional melting. 
In that case, investigating the porous target may correspond to the additional methane release as a result of crushing the clathrate cage. 
We use this same EOS for the ice impactor and the water-ice layers in the target.
To explore the effect of impactor size, we consider impactor diameters of 2, 4, 8, 10, 20, and 40 km. 
These impactors are all large enough to pass through Titan's atmosphere with minimal disruption, and form a crater on the surface \cite{Artemieva:2003}.
To manage computational costs, our high-resolution zone has a cell size of 50 m for 2 and 4-km-diameter impactors, 200 m for 8 and 10-km-diameter impactors, and 500 m for 20 and 40-km-diameter impactors, respectively.
To maintain a minimum resolution of 10 cells for our 5 km thick clathrate layer, the minimum spatial resolution we consider is a cell size of 500 m. We use the same spatial resolution for the 2 and 4, 8 and 10, and 20 and 40 km diameter impactor simulations. 
This approach of keeping spatial resolution more consistent across runs results in variation in resolution in terms of cells per projectile radius (CPPR) such that $D_{\rm imp}$ = [2, 4, 8, 10, 20, 40] corresponds to CPPR = [20, 40, 20, 25, 20, 40]. 
A similar approach was used by \citeA{Johnson:2018a} with resolution ranging from 40--70 CPPR while spatial resolution was held constant. 
We also note that our minimum resolution of 20 CPPR is sufficient to resolve the heated materials in targets \cite<e.g.,>{Pierazzo:1997,Davison:2014,Wakita:2019a}.
While we use a total ice thickness of 60 km for 2 and 4-km-diameter impactors, we include an ocean for impactors larger than 8 km with a total ice thickness of 100 km.
Note that the ocean thickness is 20 km for 200 m resolution cases and 120 km for 500 m ones. 
The ice shell’s total thickness or the presence of the ocean affects the morphology of the crater \cite<e.g.,>{Crosta:2021}. 
However, our focus is on the released methane and our results are insensitive to those settings.
Note that we use ANEOS for water \cite{Turtle:2001} for the subsurface ocean.

To estimate the mass of methane released by an individual impact, we use Lagrangian tracer particles in iSALE-2D. 
These tracer particles are emplaced in each cell before the impact and record the pressure and temperature that a parcel of material experiences during the impact.
Methane clathrate is stable when its pressure and temperature are below the dissociation curve of methane clathrate \cite{Sloan:2007,Levi:2014}. 
If methane clathrate crosses the dissociation curve, however, it will dissociate and release methane gas.   
Here we assume that methane clathrate releases all methane gas in a given cell when the tracer particle's temperature and pressure exceeds the dissociation curve.
We track the pressure and temperature of tracer particles every second from the beginning of the simulation until 600 seconds after the impact. 
We assume methane is instantaneously released from the parcel of material that a tracer tracks, when the pressure and temperature exceed the dissociation curve of methane clathrate. 
Note that we might miss some particles that exceed the curve during the one-second output interval. 
Nevertheless, the continuous distribution of material from which methane is released (see Section \ref{sec:results}) indicates our temporal resolution is sufficient.
We calculate the volume of methane clathrate based on the initial spacing and locations of tracer particles \cite{Johnson:2014}. 
To calculate the mass of released methane, we assume a methane-clathrate density of 940 kg/m$^3$ \cite{Sloan:2007} and the typical structure of methane clathrate (1 mol of methane per 5.75 mol of pure water ice \cite{Speight:2019}). 

Next, to estimate the semi-steady state rate at which impacts release methane, we couple our results for the amount of methane released by a single impact with estimates of the size-dependent impact rate at Titan for impactor diameters larger than 2 km. 
We consider a widely used impact rate \cite<Case A,>{Zahnle:2003}, which is $5.4 \times 10^{-5}$ of the annual impact rate of comets on Jupiter (i.e., 0.005 per year). 
Note that \citeA{Nesvorny:2023} reported a 20\% lower rate on Titan ($4.5 \times 10^{-5}$) by considering destruction of comets, which is based on the physical lifetime of (active) comets \cite{Nesvorny:2017}. 
In this work we use the results of \citeA{Zahnle:2003}; using the result of \citeA{Nesvorny:2023} would lower the production rate of methane released via impacts by $\sim$20\%.
We then compare the production rate of our model to the net loss rate of methane via photolysis \cite{Yung:1984, Wilson:2004}.

Finally, we explore the role that impacts might play in transitioning Titan from a cold, methane-free atmosphere to the warm, methane-rich present-day atmosphere \cite{Lorenz:1997}. 
Previous research has explored how much energy is deposited directly into the atmosphere by the impact \cite{Zahnle:2014}. 
Here, we investigate the amount of warming that could occur from impact-released methane. 
The presence of atmospheric methane will lead to the trapping of outgoing longwave radiation from Titan's surface and a greenhouse effect. 
If enough methane is released into the atmosphere a positive feedback will occur, resulting in the mass movement of methane from liquid surface reservoirs to the atmosphere. 
This, in turn, could result in further warming, and ultimately produce conditions similar to the current day \cite{Lorenz:1997}.

In this calculation, we assume that the impact-derived methane is the first introduction of methane in Titan's nitrogen atmosphere; that is there is no methane in the atmosphere prior to the methane released from impacts. 
With no methane in Titan's atmosphere, and no corresponding greenhouse effect, the equilibrium surface temperature of Titan would be 80 K \cite{McKay:1991}.
The current methane mass in the atmosphere is $2\times10^{17}$ kg and the current surface temperature is 94 K \cite{Tobie:2012}. 
The photodissociation of methane in the upper atmosphere leads to the formation of haze, which has an anti-greenhouse effect and reduces Titan's surface temperature by 9 K.
Thus, the surface temperature without any atmospheric haze would be $\sim~$105 K \cite{McKay:1991}. 

As a first-order approximation, rather than utilizing a complete radiative convective model, we assume a simple linear relationship between the surface temperature of Titan estimated from radiative equilibrium end points.  
That is the surface temperature of Titan's no atmospheric methane state (80 K, 0 kg of methane) and the surface of Titan with current day atmospheric methane but no anti-greenhouse effect from upper atmospheric haze (105 K, $2\times10^{17}$ kg of methane). 
This relationship suggests that Titan's surface temperature will increase by roughly 1 K per $\sim 10^{16}$ kg of methane released by impacts into the atmosphere.

The model proposed above also assumes that once methane is introduced, the production of haze does not result in a strong anti-greenhouse effect on the time scale it takes the nitrogen atmosphere to respond to atmospheric heating from methane emplacement. 
Using the current day mass of nitrogen in Titan’s atmosphere, the solar irradiance incident on Titan, and the heat capacity of atmospheric nitrogen it would take 23--32 years to heat the nitrogen from 80 to 105 K, our simple model end points (see \ref{app:haze}). 
In that time frame, photodissociation of methane would produce a fractal haze layer with an optical depth ranging from 0.08 to 0.23 based on current day haze particle properties \cite{Tomasko:2005} and production rates \cite{Lavvas:2008}, cross sectional area estimates for fractal haze particles \cite{Tazaki:2021a,Tomasko:2005}, and scattering efficiencies for fractal haze \cite{Es-sayeh:2023}.
Further details on these calculations are provided in \ref{app:haze}.
%If the impacts studied here release less than the current-day methane mass into Titan’s atmosphere, there would be less methane available to photodissociate and recombine into haze, the nitrogen atmosphere would take less time to heat up from greenhouse warming due to the presence of methane, and the optical thickness of fractal haze would be reduced further still. 
As such we conclude that the anti-greenhouse effects of haze production can be ignored within our simple model, so long as we only consider the immediate impacts of methane release and greenhouse warming. 

\section{Results} \label{sec:results}
\begin{figure}
\noindent\includegraphics[width=\textwidth]{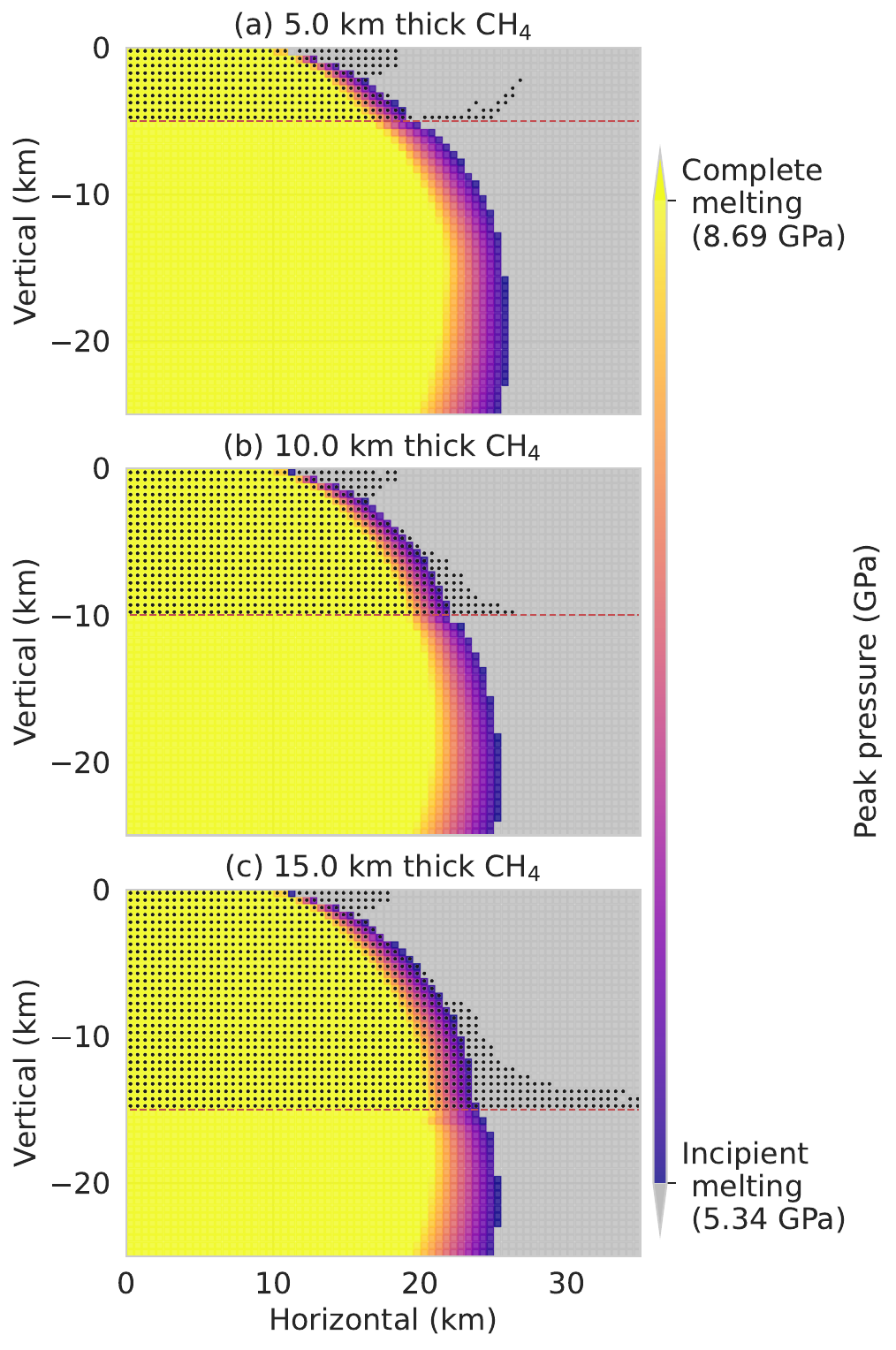}
\caption{Provenance plot of released methane by 20-km-diameter impactors.
Black dots indicate provenance of released methane; tracer particles plotted at their original pre-impact locations. 
Color illustrates their peak pressure; un-melted locations are shown in gray color.
Red dashed lines depict material boundaries.
}
\label{fig:dimp20km}
\end{figure}

Our simulations indicate that impacts into Titan's crust will release methane gas from the methane-clathrate layer. 
Figure \ref{fig:dimp20km} shows provenance plots of methane released from the target during an impact. 
The figure also shows provenance of peak pressures focused on melted target material \cite<note that we use the peak pressure of incipient melting to account for the melt, see>{Wakita:2023}.
Interestingly, most of the released methane comes from regions of the methane-clathrate target that melted during the impact.
There are only small regions within the target where methane is released from un-melted methane clathrates, i.e., regions where the temperature exceeds the dissociation curve while the pressure remains below the peak pressure for incipient melting \cite<5.34 GPa, see>{Wakita:2023}.

\begin{figure}
\noindent\includegraphics[width=\textwidth]{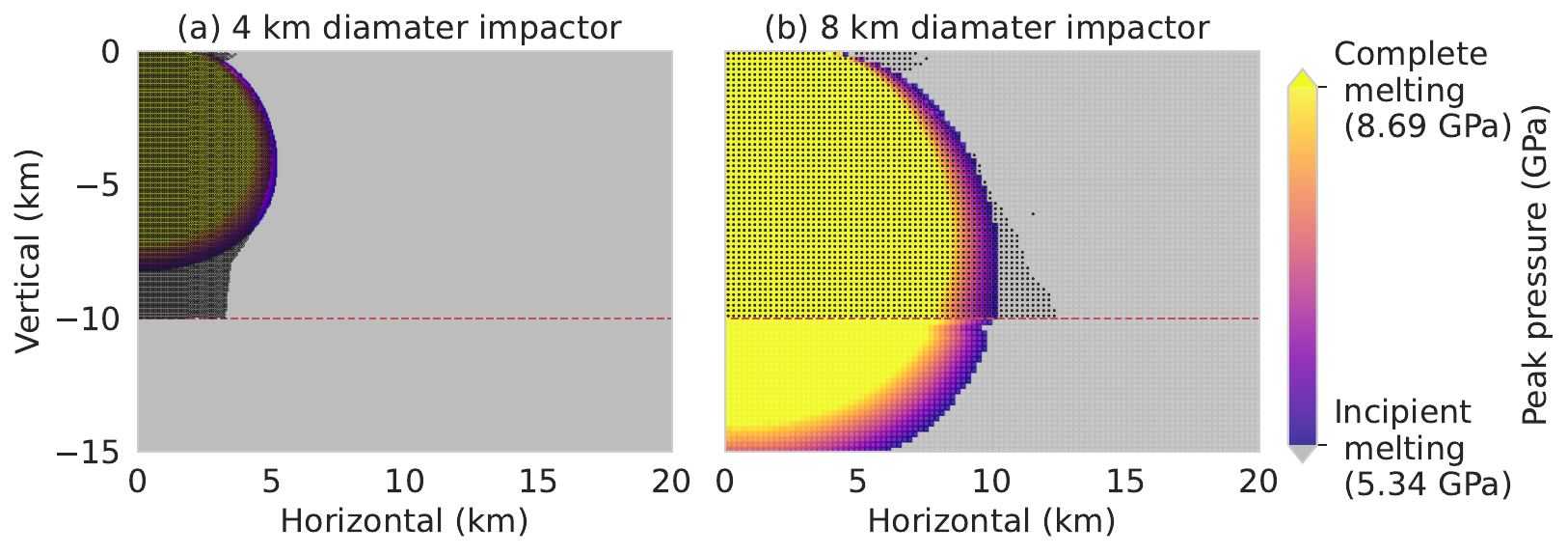}
\caption{Provenance plots of released methane similar to those in Figure \ref{fig:dimp20km}, but for different impactor sizes into a 10-km-thick methane-clathrate layer. 
Panel (a) is a 4-km-diameter impactor and (b) is an 8-km-diameter impactor.
}
\label{fig:hc10}
\end{figure}

This strong correlation between the location of released methane and the location of melted target rocks leads to a positive correlation between impactor size and the amount of methane released. 
Larger impacts will melt larger regions of the target \cite{Pierazzo:1997}.
Figure \ref{fig:hc10} shows provenance plots of released methane from a 10-km-thick methane-clathrate layer for a range of impactor sizes (4 and 8 km). 
This illustrates that the amount of methane released depends strongly on the size of the impactor, scaling roughly as the cube of the impactor size. 
Figure \ref{fig:ch4mass} and Table \ref{tab:ch4mass} summarize our results and indicate that smaller impactors are not capable of penetrating the clathrate layer.
Because these impactors create craters that are entirely within the clathrate layer, the amount of methane released does not depend strongly on the thickness of the clathrate layer (see Table \ref{tab:ch4mass}).

As the impactor becomes larger, however, the impact heats the crust through the methane-clathrate layer and into the underlying water ice layer.
While the threshold depends on the clathrate-layer thickness, impactors larger than 8 km in diameter are capable of heating the crust through even the thickest methane-clathrate layers considered in this study (Figures \ref{fig:dimp20km}, \ref{fig:hc10}(b), and \ref{fig:ch4mass}).
In this regime, the amount of methane released is less sensitive to the impactor size, scaling roughly as the square of the impactor size. Instead, for these craters, the thickness of the methane-clathrate layer dominates the amount of methane released by an impact.
A 10-km-diameter impactor into a 15-km-thick methane-clathrate layer releases up to 5 times more methane than an impact into a 5-km-thick layer (see Table \ref{tab:ch4mass}).

The amount of methane released by an impact also depends on the impact angle. 
Previous work showed that oblique impacts heat shallower but wider regions in the target than vertical impacts \cite{Pierazzo:2000a,Pierazzo:2000b,Davison:2011,Davison:2014,Wakita:2019a,Wakita:2022b}.
To investigate the effect of impact angle, we perform two oblique impact simulations using iSALE-3D \cite{Hirt:1974,Elbeshausen:2009,Elbeshausen:2011}.
A 20 km-diameter impactor striking a 10-km-thick methane-clathrate layer with an impact angle of $45^\circ$ releases 2.3 times more methane than a vertical impact (see Figure \ref{fig:3d}). 
This is not a consequence of using iSALE-3D versus iSALE-2D -- a vertical impact with the same conditions in iSALE-2D and iSALE-3D produces the same amount of released methane (within 10\%), even with different resolutions (500 m for iSALE-2D and 1000 m for iSALE-3D). 
iSALE-3D simulations of \citeA{Davison:2014} found that simulations at 10 CPPR produce appoximately 10\% less heated material compared to simulations at 20 CPPR.
The following work indicated that considering shear heating produces more heated material (i.e., melt) and resolution dependence is greatly reduced \cite{Wakita:2019a}. 
Therefore, the slightly improved accuracy is not worth the computational expense in this work (see Section \ref{sec:discussions}). 
While there could be a variation due to the target material (i.e., rock or ice), the error of $\sim$10 \% is similar between previous work and ours.
Since oblique impacts are more frequent than vertical impacts \cite{Shoemaker:1962}, we expect the total released methane via impacts on Titan is more than predicted by our vertical impact results. 
Note that we chose not to run all of our impact simulations in iSALE-3D, instead focusing on vertical impacts with iSALE-2D. 
The necessity of exploring the influence of impact angles would also require substantial computational costs, thus more oblique impact simulations are unfeasible.

Porosity in the crust also enhances the release of methane gas. 
Porous targets produce more melt than nonporous targets, because the compaction of porosity in the target increases the internal energy leading to additional heating \cite{Wunnemann:2008}.
Therefore, we expect a porous methane-clathrate layer to release more methane than a nonporous target. 
While there is no observational data on the porosity of Titan's surface, we consider 10\% and 20\% porosity in the 10-km-thick methane-clathrate layer as reasonable cases to explore. 
Note that we assume no porosity in the warm water-ice layer underlying the clathrate layer, because viscous pore closure should remove all porosity in this weak water-ice layer \cite{Fowler:1985,Johnson:2017}.
Compared to nonporous methane clathrate, 20\% porous targets release 1.4 and 1.3 times more methane mass by 4 and 20 km-diameter impactors, respectively (see Figures \ref{fig:porosity20km} and \ref{fig:porosity4km}).
In the case of 10\% porosity, methane mass increases by 1.2 and 1.1 times for 4 and 20 km-diameter impactors, respectively. 

Finally, we consider liquid methane within the voids of a porous layer, (analogous to groundwater on Earth) as a potential source of additional methane gas. 
To evaluate the transition from liquid methane to gaseous methane, we compare the specific entropies using the methane pressure--entropy phase diagram \cite{Lemonn:2010} and the corresponding pressures on the Hugoniot curve \cite<e.g.,>{Pierazzo:1997}.
\citeA{Tan:2022} indicated the liquid methane could present 5 km from the surface if there is a 15-km-thick methane-clathrate layer. 
Given the temperature of $\sim$150 K at a depth of 5 km, the corresponding depth would be 3.5 km for a 10-km-thick methane-clathrate layer. 
As a test case, we consider a 10-km-thick methane-clathrate layer that has a 3.5-km-thick liquid-saturated layer with a crust that has 10\% porosity. 
Note that liquid methane can absorb atmospheric nitrogen; at the temperatures and pressures relevant to a methane-clathrate capped crust, this liquid will be $\sim$50:50 methane--nitrogen and will have an average density of 525 kg/m$^3$ \cite{Tan:2022}.
We assume the liquid methane will have a limited effect on the impact dynamics and simply use the entropy and the pressure of tracer particles from our simulations of a 4-km-diameter impactor hitting a 10-km-thick methane-clathrate layer to calculate the volume of vaporized methane. 
We find that the additional methane released from the liquid-saturated layer is 7\% more than the non-saturated layer case. 
Vaporizing this mass of methane requires $\sim$0.1\% of the impact's energy, validating our assumption that the liquid methane will have a minimal influence on the cratering process. 
Note that even when the impact converts the methane from liquid to gas, some liquid methane can still remain within the liquid-saturated layer.
For a 20-km-diameter impactor case, it produces 4\% more additional methane.
Because the near-surface thermal gradient varies inversely with the methane-clathrate layer thickness \cite{Kalousova:2020}, the depth to which liquid methane is stable should depend directly on the methane-clathrate-layer thickness. 
Note that the variation in CPPR between 4-km-diameter impactor (CPPR=40) and 20-km-diameter impactor (CPPR=20) might also cause a small difference in the amount of additional methane released.
Nevertheless, the relative amount of methane liberated from liquid methane is similar regardless of clathrate thickness, since the released methane mass depends on the clathrate thickness in the case of larger impactors (see Figure \ref{fig:ch4mass}).
Overall, the additional methane gas released from the liquid-saturated layer is only a few percent, and its effect appears to be minimal.

\section{Discussion} \label{sec:discussions}
\begin{figure}
\noindent\includegraphics[width=0.5\textwidth]{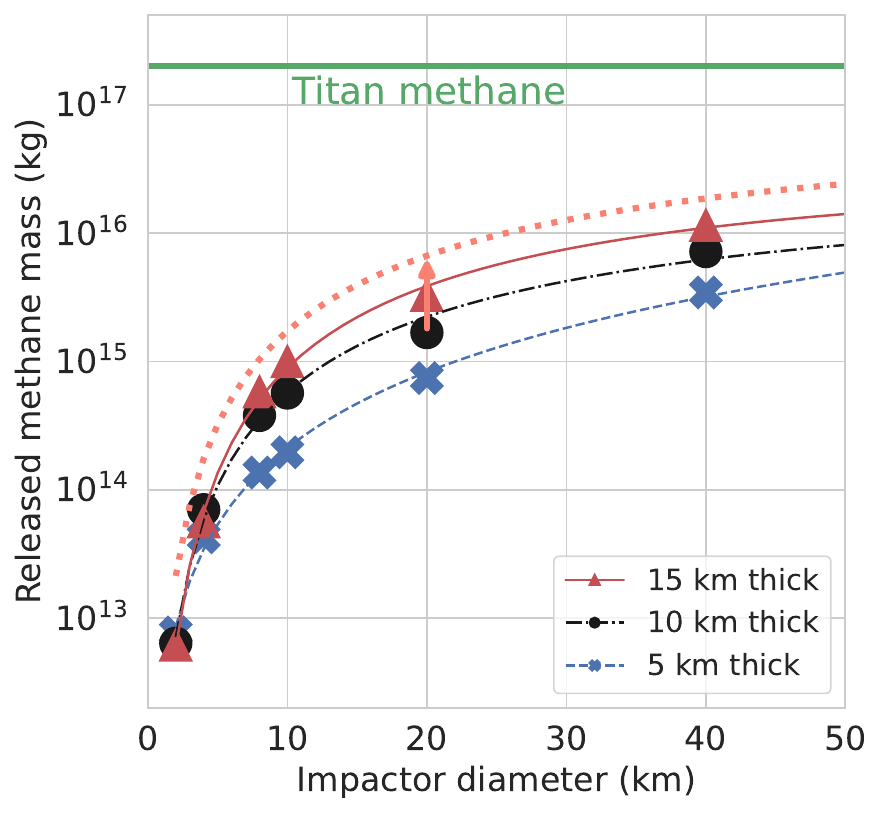}
\caption{Released methane mass as a function of impactor diameter for different methane-clathrate layer thicknesses. 
Each symbol shows released methane from a single impact into a different thickness of methane clathrate. 
The horizontal green line represents the current methane mass in Titan’s atmosphere \cite<$2\times10^{17}$ kg, >{Tobie:2012}. 
Colored lines are fitted lines of our results (see Table \ref{tab:equation}).
An orange arrow with a corresponding dotted line indicates the additional mass that could be released due to oblique impacts and porosity (see text).}
\label{fig:ch4mass}
\end{figure}

\begin{figure}
\noindent\includegraphics[width=\textwidth]{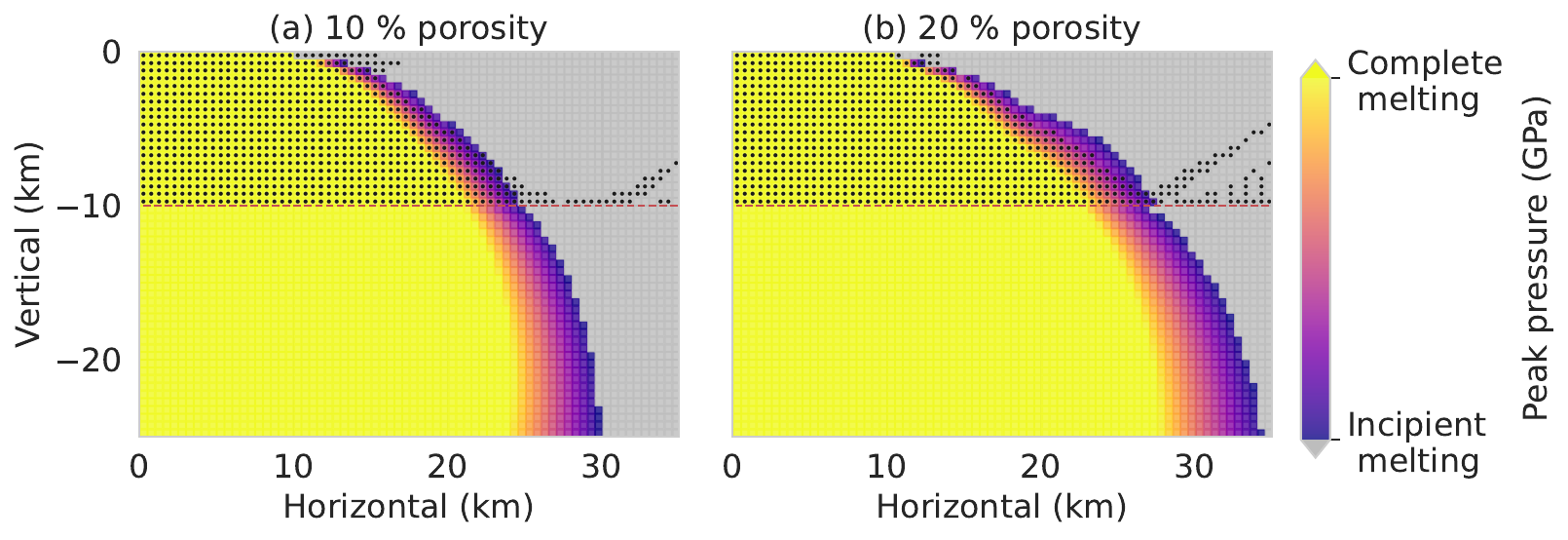}
\caption{Provenance plots of released methane similar to those in Figure \ref{fig:dimp20km}(b), but for different porosities.
Panel (a) illustrates the case of 10\% porosity and (b) illustrates the case of 20\% porosity.
Note that the color is scaled to each case, i.e., incipient melting of 3.45 GPa and 2.31 GPa and complete melting of 5.92 GPa and 4.12 GPa for 10\% and 20\% porosity, respectively.
}
\label{fig:porosity20km}
\end{figure}

\begin{figure}
\noindent\includegraphics[width=\textwidth]{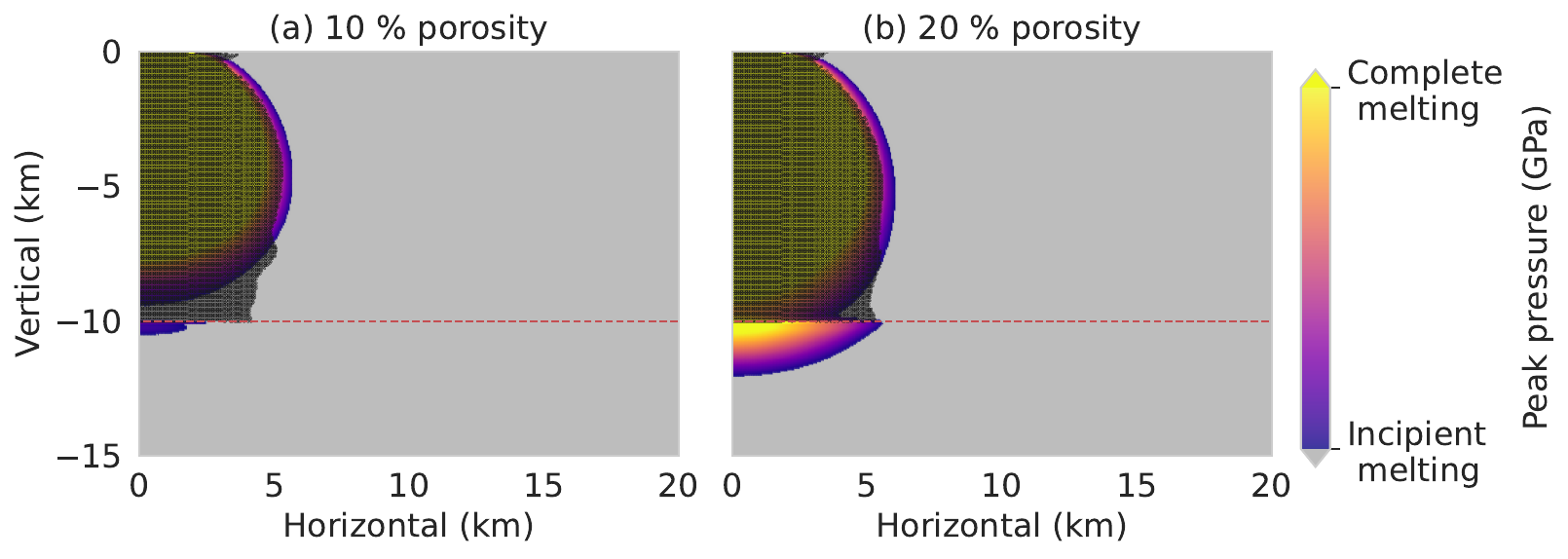}
\caption{Provenance plots of released methane similar to those in Figure \ref{fig:porosity20km}, but for a 4-km-diameter impactor.
}
\label{fig:porosity4km}
\end{figure}

\begin{figure}
\noindent\includegraphics[width=\textwidth]{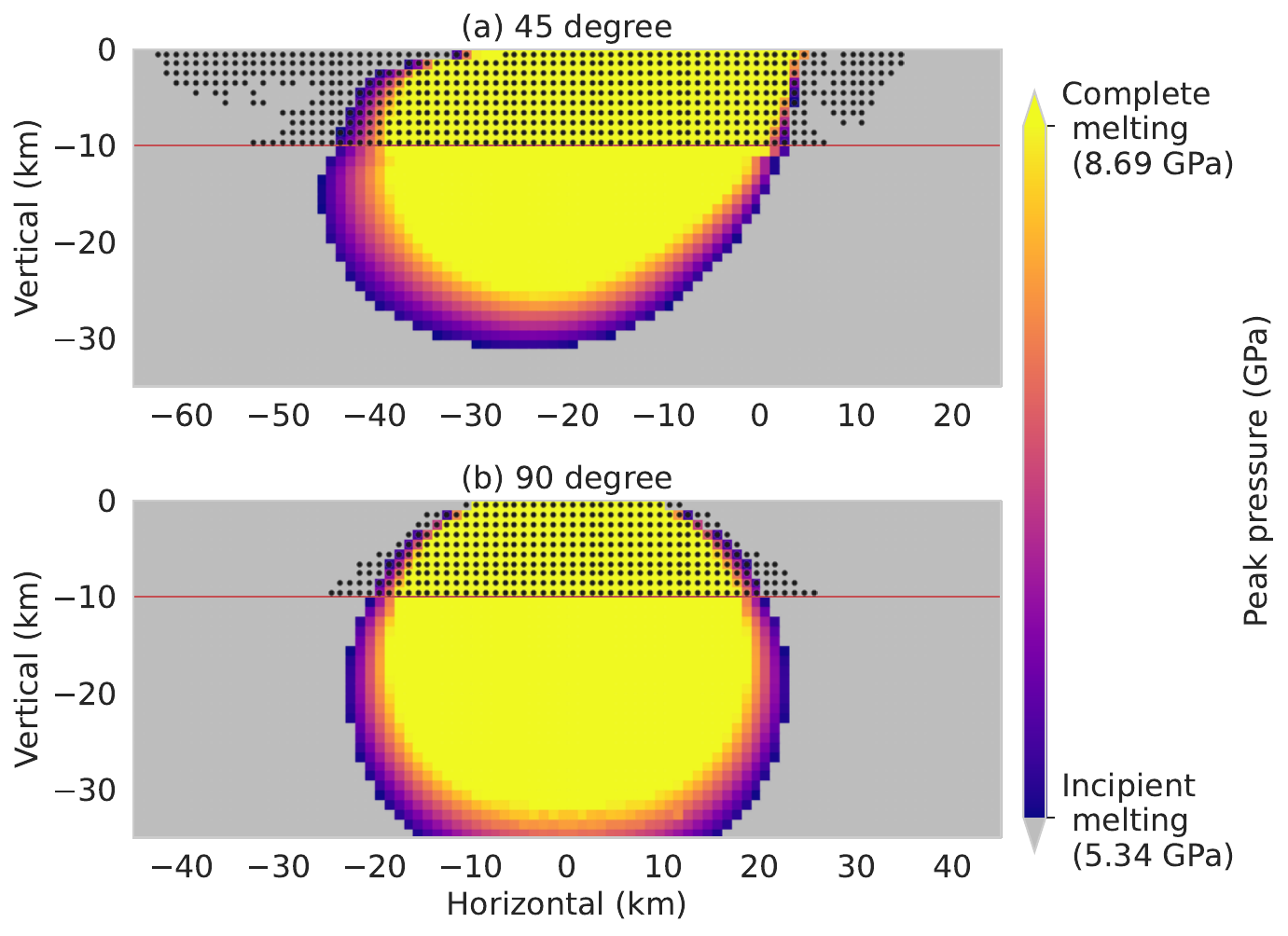}
\caption{Provenance plots of released methane into a porous target like those shown in Figures \ref{fig:dimp20km}, but using iSALE-3D. 
Panel (a) represents an impact angle of 45$^\circ$ and (b) represents an impact angle of 90$^\circ$ (vertical impactor).
Note that the impact site is the origin (0,0). 
}
\label{fig:3d}
\end{figure}

\begin{figure}
\noindent\includegraphics[width=0.5\textwidth]{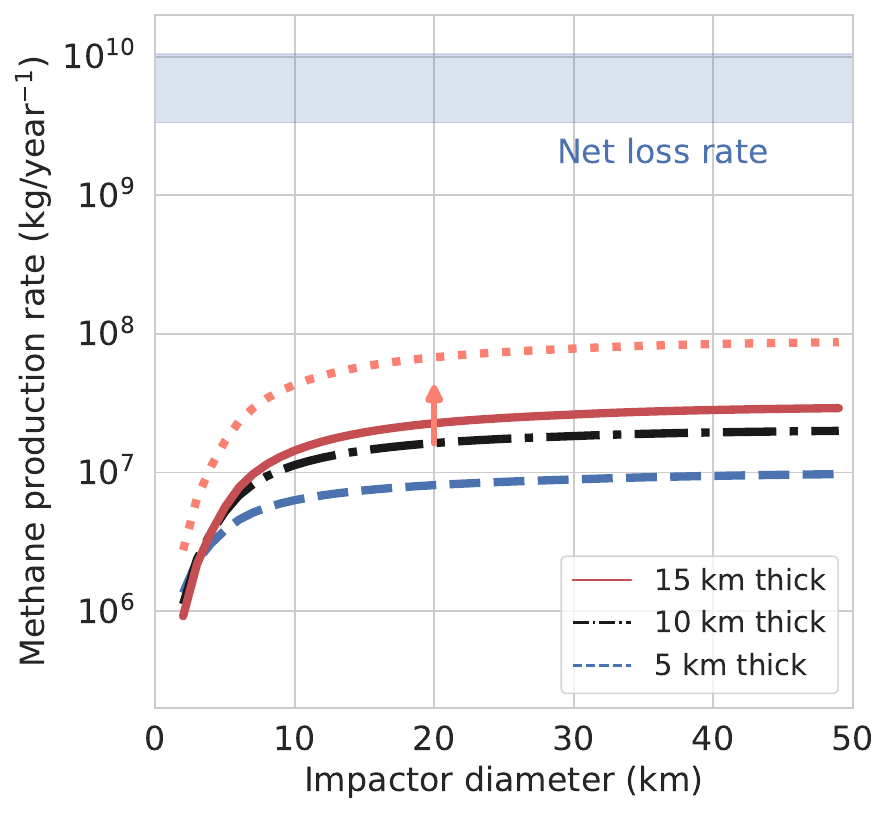}
\caption{Cumulative methane production rate as a function of impactor size. 
Each colored line represents our fitted lines for different thicknesses of methane clathrate (see legend). 
The blue shaded region illustrates the net loss rate due to photochemistry \cite{Yung:1984,Wilson:2004}.
A light red arrow with a corresponding dotted line indicates the additional mass due to oblique impacts and increased porosity (see text).}
\label{fig:ch4rate}
\end{figure}

We confirm that impacts into a methane-clathrate layer release methane gas into Titan's atmosphere. 
We find, for example, that a single 20-km-diameter impactor striking a 10-km-thick methane-clathrate layer releases $1.7\times10^{15}$ kg methane, or $\sim$0.8\% of the current mass of methane in Titan's atmosphere.
We derive approximate values for the amount of released methane mass for different impactor sizes, as indicated by the colored lines in Figure \ref{fig:ch4mass} (see Table \ref{tab:equation} for details).
Using the typical impact rate on Titan \cite{Zahnle:2003}, we compare the production rate of released methane with the net loss rate due to photolysis \cite{Yung:1984,Wilson:2004}.
The production rate by impacts is 1--2 orders of magnitude lower than the net loss rate by photolysis (Figure \ref{fig:ch4rate}). 
While factors such as impact angle and porosity will increase this production rate by a factor of 2--3 (see dotted line in Figure \ref{fig:ch4mass}), we find that impacts can provide only $\sim$1--3\% of the methane needed to maintain Titan’s current methane-rich atmosphere.
Given $\sim$3\% of the current methane mass released via a single impact, it would last for $\sim$0.6 Myr under the loss rate of $\simeq$300 kg/s \cite{Wilson:2009}.
Thus, resupplying methane via impacts can enhance the lifetime of methane gas by up to 3\%.

Many small impacts do not seem capable of releasing all of the methane in Titan's atmosphere, but is it possible that one large impact could account for its presence there? 
For example, Titan's largest crater, Menrva looks morphologically quite fresh \cite{Wood:2010}. 
Could this impact have released the methane we currently see in Titan's atmosphere, and in the process, kick-started the exogenic processes responsible for its young surface age of 200 Myr to 1 Gyr \cite{Neish:2012}?
Here we estimate the amount of methane released by the formation of Menrva.
Using the crater scaling laws for porous target \cite<Eq. (5) in >{Johnson:2016c}, we find that a 40-km-diameter impactor is suitable for the formation of Menrva-size crater.
This impactor ($45^{\circ}$ impact angle) obliquely hitting a 15-km-thick with 20 \% porosity would release about $0.3 \times 10^{17}$ kg methane gas. 
Thus, the formation of Menrva provides $\sim$15\% of the current methane mass; it cannot account for the presence of Titan's methane-rich atmosphere. 
This methane contribution would only last for 3.2 Myr at the net mass loss rate of $\simeq 300$ kg/s \cite{Wilson:2009}, so its contribution has been already lost to space.

While a single impact may not release enough methane to account for the amount observed in Titan's atmosphere, methane released by impacts would contribute to the warming of Titan's surface and atmosphere through the greenhouse effect. 
We estimated previously that methane warms Titan's atmosphere by 1 K for every $\sim 10^{16}$ kg of methane released (see Section \ref{sec:methods}). 
It follows that the amount of methane released by a Menrva-forming impact, releasing $0.3 \times 10^{17}$ kg of methane as described above, would warm Titan by $\sim$3 K depending on impactor size, obliquity, and porosity. 

It is important to note, however, that impacts
can also increase the atmospheric temperature through other effects, such as the impact-induced shock wave and interaction of impact ejecta with the atmosphere \cite<e.g.,>{Melosh:1989,Zahnle:2014,Morgan:2022}. 
Assuming that half of the impact energy ultimately results in atmospheric heating, \citeA{Zahnle:2014} find that a Menrva forming impact increases the temperature of Titan by 80 K; 1.5 orders of magnitude more than the heating based on the greenhouse effect of impact-released methane. 
Thus, the temperature increase determined in this paper is insignificant and cannot be directly compared to that of \citeA{Zahnle:2014}. 
It is worth noting that these effects are expected to be minor for all crater forming impacts on Titan smaller than Menrva \cite{Zahnle:2014}.

Aside from ongoing heliocentric impacts by comets, Centaurs, and Kuiper belt objects, additional planetocentric impacts could also contribute to the impact history on Titan \cite{Dones:2009,Kirchoff:2018}. 
Because it is statistically very unlikely that there were multiple heliocentric Menrva-sized forming impacts
over the age of the solar system due to heliocentric impacts \cite<e.g.,>{Neish:2012}, multiple Menrva-sized impacts must have been planetocentric.
Note that planetocentric impactors have lower impact velocities compared to heliocentric impactors, thus the released mass would be lower than our estimates (see the following section).
One source for planetocentric impacts is ejecta from other Saturnian satellites \cite<e.g., Hyperion, >{Dobrovolskis:2004} and another may be the formation of the Saturnian ring system.
Two recent works are arguing for the formation of the Saturnian ring via disruption of one or more icy moons in the Saturnian system \cite{Wisdom:2022,Teodoro:2023}. 
The first --- the destruction of the "Chrysalis" moon that makes the Saturnian ring --- might cause multiple large impact events on Titan \cite{Wisdom:2022}. 
The second --- collisions between preexisting icy moons ---  may also bring their fragments into the orbit of Titan \cite{Teodoro:2023}.
Menrva might be the only preserved crater from these events; the others could have been eroded or buried by sediment \cite{Neish:2013,Neish:2016}.
If a Saturnian-ring-forming event or a similar moon-destroying event occurs \cite{Wisdom:2022,Teodoro:2023}, it could produce multiple large impactors on Titan over a short period. 
These large impacts might have resurfaced Titan, and Menrva could be the most recent crater that is clearly observable. 
While other craters may not be visible at present, they might have released comparable or more methane than Menrva, potentially releasing enough methane to support Titan’s present-day climate.

Alternatively, the Xanadu region has been suggested to be an impact feature \cite{Brown:2011}; such an impact could have released enough methane on its own to warm Titan to its present-day state. 
\citeA{Brown:2011} suggested that a 60-km-diameter object could reproduce Xanadu with an impactor speed of 12 km/s and oblique impact of 45 degrees. 
Extrapolating from our data, this would have provided about $0.5 \times 10^{17}$ kg methane released from a 15-km-thick methane-clathrate target with 20\% porosity, resulting in a methane atmospheric warming of $\sim$5 K. 
As we use a lower impact velocity (10.5 km/s), this would constitute a lower limit on methane release. 
Additionally, if Hotei Regio was another large impact basin \cite{Soderblom:2009}, it could have also contributed to Titan’s climate change \cite<e.g.,>{Zahnle:2014}.
It thus seems plausible that an impact of this size could release enough methane to create Titan's present-day climate, although the hypothesis that Xanadu and Hotei are impact craters remains in question.

The variation in impact velocity influences the amount of released methane. 
Since the region of released methane closely correlates with the melt region (see Figures \ref{fig:dimp20km} and \ref{fig:hc10}), we can apply the velocity dependence of the impact melt \cite{Pierazzo:1997}.
For an ice-on-ice impact scenario, the melt volume is proportional to the impact velocity to the power of $\sim$1.3. 
Although we assume an impact velocity of 10.5 km/s in this study, we explore higher and lower impact velocities of 15 km/s and 5 km/s as test cases.
If the impact velocity increases from 10.5 km/s to 15 km/s, the released methane mass increases by about 1.6 times. 
In contrast, an impact velocity of 5 km/s releases 0.6 times lower methane mass.
Therefore, we need to care about the effect of impact velocity when we assess the planetocentric impacts (see the section above) or the distribution of the impact velocity. 

In this work, we neglect methane release from the water-ice layer. 
The convective water-ice layer might, however, be mixed with methane clathrates \cite{Carnahan:2022}, which would cause our results to be an underestimate.
Titan may hold $3.5\times10^{20}$ kg of methane in total, potentially originating from the core \cite{Miller:2019}.
The outgassed methane from the core could ascend up to the convecting water-ice layer and eventually form methane clathrates near the surface. 
If we take a 100 km total thick ice layer capped with a 10-km-thick methane-clathrate layer that possesses $10^{20}$ kg methane, the 90-km-thick water-ice layer might be mixed with $2.5\times10^{20}$ kg methane clathrate. 
Note that the assumed total methane mass of $10^{20}$ kg is three orders of magnitude larger than the current methane mass in Titan's atmosphere ($10^{17}$ kg).
Since the methane dissociation process from the mixed layer is unclear, we estimate its contribution only from the melt region for simplicity.
When a 20-km-diameter impactor hits a 10-km-thick methane-clathrate layer over a water-ice layer mixed with methane clathrate, it would release an additional 10\% of methane by mass.
As we only consider the melt region in this simple calculation, this amount might be an underestimate. 
Future work considering this and other effects (see Section \ref{sec:results}) in more detail might reveal a single large impact condition that could have triggered the greenhouse effect on Titan.

Lastly, we would like to mention the effect of resolution. To resolve the methane clathrate layer reasonably, we vary the resolution from 50 m to 500 m based on the impactor diameter (see Section \ref{sec:methods}). 
This corresponds to resolution of 20--40 cells per projectile radius (CPPR). 
Previous work suggested that a 20 CPPR is sufficient to resolve impact heating  \cite<e.g.,>{Pierazzo:1997}, and this resolution may result in $\sim$10\% error compared to simulations at 40 CPPR. 
While we take 20 CPPR as the lowest resolution, we also use 25 CPPR for 10 km, 40 CPPR for 4 km and 40 km diameter impactors. 
We performed runs at 20 CPPR for all simulations to test this effect and found $\sim\pm$10\% methane released from the simulation at 40 CPPR. 
These variations were not consistently toward less methane released at lower resolution and the results from 20 CPPR runs. 
These small variations are nearly imperceptible when plotted similarly to Figure \ref{fig:ch4mass} giving us confidence that our results are not strongly affected by our choice of resolution (see Figure \ref{fig:cppr}).
Using fits shown in Fig \ref{fig:cppr}, the difference between the estimated released methane masses of 20 km-diameter-impactors is within $\sim$3\%. 

\section{Conclusions}
We determine the amount of methane gas released via impacts into a methane-clathrate layer on Titan. 
This is a potential source to replenish methane in Titan’s atmosphere, which should be unstable over solar system timescales. 
We perform impact simulations considering 5--15-km-thick methane-clathrate layers, and a range of impactor diameters (2--40 km). 
We find that the released methane from a single impact is unlikely to be a substantial current source of methane in Titan’s atmosphere. 
Our results show that the amount of methane released correlates strongly with the size of the melt pool produced in the impact. 
The amount of released methane thus depends on the impactor size and the methane-clathrate-layer thickness. 
We also find that oblique impacts into a porous layer increase the released methane mass by 2--3 times, compared to a vertical impact into non-porous clathrate.
Although the total methane loss rate due to photolysis is greater than the methane release rate by impacts, we find that the released methane can enhance the lifetime of methane gas in Titan's atmosphere by up to 3\%.
A Menrva-sized crater formed by an oblique impact could contribute $\sim$15\% of the current methane mass, which is enough methane to warm Titan by 3 K.
Thus, if several of these impacts struck Titan, they could measurably alter Titan's climate. 
One potential source for such large impactors could be the disruption of icy moons within the Saturnian system. 

%%

%  Numbered lines in equations:
%  To add line numbers to lines in equations,
%  \begin{linenomath*}
%  \begin{equation}
%  \end{equation}
%  \end{linenomath*}

\clearpage

%% Enter Figures and Tables near as possible to where they are first mentioned:
%
% DO NOT USE \psfrag or \subfigure commands.
%
% Figure captions go below the figure.
% Table titles go above tables;  other caption information
%  should be placed in last line of the table, using
% \multicolumn2l{$^a$ This is a table note.}
%
%----------------
% EXAMPLE FIGURES
%
% \begin{figure}
% \includegraphics{example.png}
% \caption{caption}
% \end{figure}
%
% Giving latex a width will help it to scale the figure properly. A simple trick is to use \textwidth. Try this if large figures run off the side of the page.
% \begin{figure}
% \noindent\includegraphics[width=\textwidth]{anothersample.png}
%\caption{caption}
%\label{pngfiguresample}
%\end{figure}
%
%
% If you get an error about an unknown bounding box, try specifying the width and height of the figure with the natwidth and natheight options. This is common when trying to add a PDF figure without pdflatex.
% \begin{figure}
% \noindent\includegraphics[natwidth=800px,natheight=600px]{samplefigure.pdf}
%\caption{caption}
%\label{pdffiguresample}
%\end{figure}
%
%
% PDFLatex does not seem to be able to process EPS figures. You may want to try the epstopdf package.
%

%
% ---------------
% EXAMPLE TABLE
% Please do NOT include vertical lines in tables
% if the paper is accepted, Wiley will replace vertical lines with white space
% the CLS file modifies table padding and vertical lines may not display well
%
 \begin{table}
 \caption{Released methane mass for different scenarios considered in this work.}
 \label{tab:ch4mass}
 \centering
 \begin{tabular}{l c r}
 \hline
  Impactor & Methane-clathrate  & Released methane\\
  diameter (km) & thickness (km) & mass (kg) \\
 \hline
   2  & 5  & $7.75\times10^{12}$\\
   2  & 10  & $6.38\times10^{12}$   \\
   2  & 15  &  $6.18\times10^{12}$  \\
   4  & 5  & $4.28\times10^{13}$\\
   4  & 10  & $7.01\times10^{13}$   \\
   4  & 15  &  $5.64\times10^{13}$  \\
   8  & 5  & $1.36\times10^{14}$\\
   8  & 10  & $3.80\times10^{14}$   \\
   8  & 15  &  $5.80\times10^{14}$  \\
   10  & 5  & $1.96\times10^{14}$\\
   10  & 10  & $5.68\times10^{14}$   \\
   10  & 15  &  $9.99\times10^{14}$  \\
   20  & 5  & $7.42\times10^{14}$\\
   20  & 10  & $1.68\times10^{15}$   \\
   20  & 15  &  $3.28\times10^{15}$  \\
   40  & 5  & $3.44\times10^{15}$\\
   40  & 10  & $7.17\times10^{15}$   \\
   40  & 15  &  $1.15\times10^{16}$  \\
 \hline
 %\multicolumn{2}{l}{$^{a}$Footnote text here.}
 \end{tabular}
 \end{table}

 \begin{table}
 \caption{Equation coefficients of released methane mass ($y$)$^{a}$}
 \label{tab:equation}
 \centering
 \begin{tabular}{l c c c}
 \hline
  Methane-clathrate & A & B & C \\
  thickness (km) &  &  &  \\
 \hline
   5 & -0.0241  & 2.02 & 12.3 \\
   10 & -0.776  & 3.73 & 11.7   \\
   15 & -1.00  & 4.43 & 11.5  \\
 \hline
 \multicolumn{3}{l}{$^{a}$ \(log_{10}(y) = Alog_{10}(x)^2+Blog_{10}(x)+C\), where $x$ is an impactor diameter.}
 \end{tabular}
 \end{table}

%% SIDEWAYS FIGURE and TABLE
% AGU prefers the use of {sidewaystable} over {landscapetable} as it causes fewer problems.
%
% \begin{sidewaysfigure}
% \includegraphics[width=20pc]{figsamp}
% \caption{caption here}
% \label{newfig}
% \end{sidewaysfigure}
%
%  \begin{sidewaystable}
%  \caption{Caption here}
% \label{tab:signif_gap_clos}
%  \begin{tabular}{ccc}
% one&two&three\\
% four&five&six
%  \end{tabular}
%  \end{sidewaystable}

%% If using numbered lines, please surround equations with \begin{linenomath*}...\end{linenomath*}
%\begin{linenomath*}
%\begin{equation}
%y|{f} \sim g(m, \sigma),
%\end{equation}
%\end{linenomath*}

%%% End of body of article

%%%%%%%%%%%%%%%%%%%%%%%%%%%%%%%%
%% Optional Appendix goes here
%
% The \appendix command resets counters and redefines section heads
%
% After typing \appendix
%
%\section{Here Is Appendix Title}
% will show
% A: Here Is Appendix Title
%
\clearpage
\appendix
\section{Atmospheric model -- Heating and anti-greenhouse considerations} \label{app:haze}
We have proposed a simple atmospheric model to explore the heating effects of methane addition to Titan’s atmosphere. 
This model assumes impact released methane is the first methane to be added to Titan’s atmosphere, the initial atmosphere is composed entirely of nitrogen, and once methane is released the formation of haze through photodissociation, and introduction of an anti-greenhouse effect, does not outpace the warming of atmospheric nitrogen. 
To validate this, we calculated the time it would take to heat a purely nitrogen atmosphere and the optical thickness of haze that would be created in that timeframe using current day nitrogen atmospheric mass, methane atmospheric mass, and observed haze particle parameters.

The specific heat capacity of nitrogen gas at constant pressure is 29.1 J/mol K at 100 K \cite{Chase:1998}. 
Using the partial pressure of nitrogen at the surface of Titan, $1.43 \times 10^5$ Pa, or 95\% of the total surface pressure of $1.50 \times 10^5$ Pa \cite{Lodders:1998}, we find a total atmospheric mass of nitrogen of $8.80 \times 10^{18}$ kg or $3.14 \times 10^{20}$ moles. 
Therefore, the amount of energy needed to raise the temperature of Titan’s atmospheric nitrogen by 1 K is $9.14 \times 10^{21}$ J.

To estimate a timeframe over which to assess the optical depth of haze formation, we found the time it would take our purely nitrogen atmosphere to heat from our equilibrium surface temperature end points -- 80 K for a no methane atmosphere to 105 K for a current day methene with no anti-greenhouse effect atmosphere \cite{McKay:1991}.
Using the solar constant at the average Sun-Titan distance of 15.04 W/m$^2$ \cite{Lodders:1998}, we find that Titan receives $3.13 \times 10^{14}$ J/s. 
Using this with the energy needed to raise the temperature by 1 K it would take 23.2 Earth years to heat the nitrogen atmosphere 25 K (from 80 to 105 K). 
If we assume that the bond albedo of Titan was the same as current day, 0.27 \cite{Younkin:1974}, the time frame for heating would increase to 31.8 Earth years. 

The total number of particles produced through methane photodissociation is a function of the production rate of haze formation and the timeframe over which production takes place:
\begin{equation}
    N_{\rm{tot}} = 3 \times 10^{-14} [\rm{g/cm^2/s}] t [\rm{s}] 1.66 \times 10^{21} [\rm{g}^{-1}], 
\end{equation}
where $3 \times 10^{-14} [\rm{g/cm^2/s}]$ is the haze production rate \cite{Lavvas:2008}, $t$ is the time in seconds to heat our nitrogen atmosphere by 25 K, and $1.66 \times 10^{21} [\rm{g}^{-1}]$ is a conversion for 1000 Dalton particles. 
Therefore, the total number of haze particles produced, $N_{\rm{tot}}$, ranges from $3.65 \times 10^{16} [\rm{cm}^{-2}]$ for 23.2 years to $4.98 \times 10^{16} [\rm{cm}^{-2}]$ for 31.8 years of production. 

We then considered the cross-sectional area of haze particles and their scattering efficiency to determine the optical depth of the haze. 
First, cross sectional area, $G$, was found using two expressions:
\begin{equation}
    G = N_{\rm{mon}} \pi R_o^2 (0.352 + 0.566 N_{\rm{mon}}^{-0.138}) [\rm{cm}^2],
\end{equation}
and 
\begin{eqnarray}    
    R_{\rm{eff}} &=& R_o \sqrt{N_{\rm{mon}}^{0.925}} [\rm{cm}], \\
    G &=& \pi R_{\rm{eff}}^2 [\rm{cm}^2],
\end{eqnarray}
from \citeA{Tazaki:2021a} and \citeA{Tomasko:2005}, respectively, where $N_{\rm{mon}}$ is the number of monomers per haze particle, 5000, and $R_o$ is the single monomer size, 50 nm, derived from \textit{in situ} haze particle parameters observed during the Huygens probe descent \cite{Tomasko:2005}. 
These provided cross sectional areas of $2.08 \times 10^{-7} \rm{cm}^2$ and $2.07 \times 10^{-7} \rm{cm}^2$, respectively. 

Next, we considered the scattering efficiency of the haze particles given their cross-sectional area and a fractal dimension of 2.4 \cite{Es-sayeh:2023}. 
Scattering efficiency, $Q_s$, ranged from $0.226 \times 10^{-10}$ at 550 nm to $0.114 \times 10^{-10}$ at 880 nm. 
The optical depth of the haze, $\tau$, can then be found as
\begin{equation}
\tau = G Q_s N_{\rm{tot}}.
\end{equation}
Putting this all together we find that haze optical depth ranges from 0.086 to 0.17 for the 23.2 year case and 0.12 to 0.23 for the 31.8 year case. 
Thus, the production of haze over a 30 year time frame does not substantially increase the optical depth of the atmosphere and we can, for the purposes of this study where less than current day methane mass is present and are considering warming of only a few Kelvin, ignore the anti-greenhouse effects of the haze. 
Further, more detailed, studies may be useful to better understand the interplay between impact released methane, haze optical depth, and anti-greenhouse effects.

We also took some time to explore the effects of haze particle size on optical depth. 
It is possible that the first haze formed after methane release would be smaller than that detected in the current day atmosphere of Titan \cite{Tomasko:2005}. 
The size of a haze particle directly corresponds to the number of monomers that make up the particle. 
For haze particles with $N_{\rm{mon}}=1000$, representing a particle size 1/5th that of the calculations discussed above, their cross sectional area reduces to $4.67 \times 10^{-8} \rm{cm}^2$ and $4.01 \times 10^{-8} \rm{cm}^2$ respectively. 
Using the same fractal dimension of 2.4 \cite{Es-sayeh:2023}, $Q_s$ becomes $0.445 \times 10^{-11}$ at 550 nm to $0.190 \times 10^{-11}$ at 880 nm. 
We then use the same haze production rate \cite{Lavvas:2008}, $N_{\rm{tot}}$, regardless of the number of monomers composing each particle. 
As a result, the optical depths range from 0.0027 to 0.0075 for 23.2 year case and 0.0037 to 0.010 for the 31.7 year case. 
For larger haze particles with $N_{\rm{mon}}=10000$, representing 2 times that of the calculations discussed above, their cross sectional area increases to $4.01 \times 10^{-7} \rm{cm}^2$ and $3.93 \times 10^{-7} \rm{cm}^2$, and $Q_s$ is $0.428 \times 10^{-10}$ at 550 nm to $0.239 \times 10^{-10}$ at 880 nm. 
Thus, the optical depths for larger haze particles range from 0.34 to 0.62 for 23.2 year case and 0.46 to 0.85 for the 31.7 year case. 
As such, the smaller haze results in a smaller optical depth while the larger haze results in a larger optical depth.
One caveat is that the haze formation rate, which we assume to be constant on the basis of the Lyman alpha photon flux, may have some dependence on the number of monomers and thus the size of the haze particles. 
Although this could have an effect on our estimates, adding this level of complexity is far beyond the focus of this paper, requiring more observations and detailed studies, and we leave this quandary to motivate future studies.

\section{Low resolution results} \label{app:low}
\begin{figure}
\noindent\includegraphics[width=\textwidth]{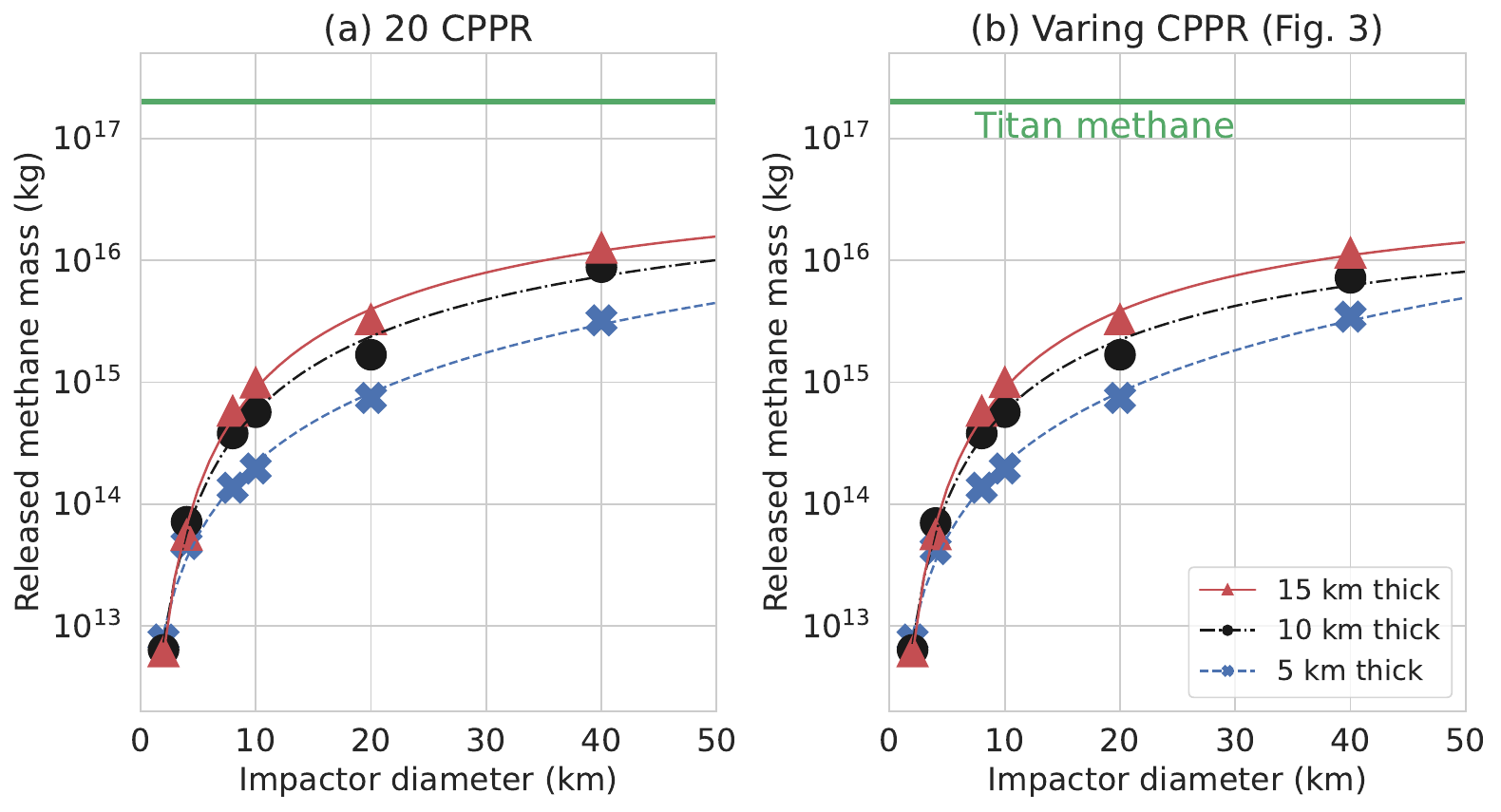}
\caption{
Released methane mass as a function of impactor diameter similar to those in Figure \ref{fig:ch4mass}, but all results using 20 CPPR (a).
Figure \ref{fig:ch4mass} is also shown in panel (b).
}
\label{fig:cppr}
\end{figure}

Figure \ref{fig:cppr} shows the results with 20 CPPR along with Figure 3.

%%%%%%%%%%%%%%%%%%%%%%%%%%%%%%%%%%%%%%%%%%%%%%%%%%%%%%%%%%%%%%%%
%
% Optional Glossary, Notation or Acronym section goes here:
%
%%%%%%%%%%%%%%
% Glossary is only allowed in Reviews of Geophysics
%  \begin{glossary}
%  \term{Term}
%   Term Definition here
%  \term{Term}
%   Term Definition here
%  \term{Term}
%   Term Definition here
%  \end{glossary}

%
%%%%%%%%%%%%%%
% Acronyms
%   \begin{acronyms}
%   \acro{Acronym}
%   Definition here
%   \acro{EMOS}
%   Ensemble model output statistics
%   \acro{ECMWF}
%   Centre for Medium-Range Weather Forecasts
%   \end{acronyms}

%
%%%%%%%%%%%%%%
% Notation
%   \begin{notation}
%   \notation{$a+b$} Notation Definition here
%   \notation{$e=mc^2$}
%   Equation in German-born physicist Albert Einstein's theory of special
%  relativity that showed that the increased relativistic mass ($m$) of a
%  body comes from the energy of motion of the body—that is, its kinetic
%  energy ($E$)—divided by the speed of light squared ($c^2$).
%   \end{notation}

\clearpage

\section{Open Research}
%AGU requires an Availability Statement for the underlying data needed to understand, evaluate, and build upon the reported research at the time of peer review and publication.

%Authors should include an Availability Statement for the software that has a significant impact on the research. Details and templates are in the Availability Statement section of the Data and Software for Authors Guidance: \url{https://www.agu.org/Publish-with-AGU/Publish/Author-Resources/Data-and-Software-for-Authors#availability}

%It is important to cite individual datasets in this section and, and they must be included in your bibliography. Please use the type field in your bibtex file to specify the type of data cited. Some options include Dataset, Software, Collection, ComputationalNotebook. Ex: 
%\\
%\begin{verbatim}
%
%@misc{https://doi.org/10.7283/633e-1497,
%  doi = {10.7283/633E-1497},
%  url = {https://www.unavco.org/data/doi/10.7283/633E-1497},
%  author = {de Zeeuw-van Dalfsen, Elske and Sleeman, Reinoud},
%  title = {KNMI Dutch Antilles GPS Network - SAB1-St_Johns_Saba_NA P.S.},
%  publisher = {UNAVCO, Inc.},
%  year = {2019},
%  type = {dataset}
%}
%
%\end{verbatim}

%For physical samples, use the IGSN persistent identifier, see the International Geo Sample Numbers section:
%\url{https://www.agu.org/Publish-with-AGU/Publish/Author-Resources/Data-and-Software-for-Authors#IGSN}
All of our results were produced using iSALE-2D and our input and output files are available in \cite{Wakita:2024a}.
Please note that iSALE-2D is distributed on a case-by-case basis to academic users in the impact community,
and the usage of the iSALE-3D code is restricted to those who have contributed to the development of iSALE-2D.
It requires registration from the iSALE webpage (https://isale-code.github.io/); computational requirements are listed there.  
%%%%%%%%%%%%%%%%%%%%%%%%%%%%%%%%%%%%%%%%%%%%%%%

\acknowledgments
%This section is optional. Include any Acknowledgments here.
%The acknowledgments should list:\\
%All funding sources related to this work from all authors\\
%Any real or perceived financial conflicts of interests for any author\\
%Other affiliations for any author that may be perceived as having a conflict of interest with respect to the results of this paper.\\
%It is also the appropriate place to thank colleagues and other contributors. AGU does not normally allow dedications.

This work was supported by NASA Cassini Data Analysis Program grant 80NSSC20K0382.
We gratefully acknowledge the developers of iSALE-2D, including Gareth Collins, Kai W\"{u}nnemann, Dirk Elbeshausen, Tom Davison, Boris Ivanov, and Jay Melosh,
and the developers of iSALE-3D, including Dirk Elbeshausen, Kai W\"{u}nnemann, Gareth Collins, and Tom Davison.
We also thank Tom Davison, the developer of the pySALEPlot tool, which we used to create some plots in this work.
This research was supported in part through computational resources provided by Information Technology at Purdue, West Lafayette, Indiana.

%% ------------------------------------------------------------------------ %%
%% References and Citations

%%%%%%%%%%%%%%%%%%%%%%%%%%%%%%%%%%%%%%%%%%%%%%%
%
% \bibliography{<name of your .bib file>} don't specify the file extension
%
% don't specify bibliographystyle

% In the References section, cite the data/software described in the Availability Statement (this includes primary and processed data used for your research). For details on data/software citation as well as examples, see the Data & Software Citation section of the Data & Software for Authors guidance
% https://www.agu.org/Publish-with-AGU/Publish/Author-Resources/Data-and-Software-for-Authors#citation

%%%%%%%%%%%%%%%%%%%%%%%%%%%%%%%%%%%%%%%%%%%%%%%

%\bibliography{references.bib}

%Reference citation instructions and examples:
%
% Please use ONLY \cite and \citeA for reference citations.
% \cite for parenthetical references
% ...as shown in recent studies (Simpson et al., 2019)
% \citeA for in-text citations
% ...Simpson et al. (2019) have shown...
%
%
%...as shown by \citeA{jskilby}.
%...as shown by \citeA{lewin76}, \citeA{carson86}, \citeA{bartoldy02}, and \citeA{rinaldi03}.
%...has been shown \cite{jskilbye}.
%...has been shown \cite{lewin76,carson86,bartoldy02,rinaldi03}.
%... \cite <i.e.>[]{lewin76,carson86,bartoldy02,rinaldi03}.
%...has been shown by \cite <e.g.,>[and others]{lewin76}.
%
% apacite uses < > for prenotes and [ ] for postnotes
% DO NOT use other cite commands (e.g., \citet, \citep, \citeyear, \citealp, etc.).
% \nocite is okay to use to add references from your Supporting Information
%

\end{document}